# Phonons and Anomalous Lattice Behaviour in $KMnAg_3(CN)_6$ and $KNiAu_3(CN)_6$: Inelastic Neutron Scattering and First-Principles Calculations

Baltej Singh[1,2], Mayanak K. Gupta[1], Ranjan Mittal[1,2], Mohamed Zbiri[3], Thoguluva R. Ravindran[4], Helmut Schober[3] and Samrath L. Chaplot[1,2]

[1]*Solid State Physics Division, Bhabha Atomic Research Centre, Mumbai, 400085, India*
[2]*Homi Bhabha National Institute, Anushaktinagar, Mumbai 400094, India*
[3]*Institut Laue-Langevin, 71 avenue des Martyrs, Grenoble Cedex 9, 38042, France*
[4]*Materials Science Group, Indira Gandhi Centre for Atomic Research, Kalpakkam 603 102, India*

Cyanide based framework compounds are known to show large negative thermal expansion behaviour. Here we report the phonon and anomalous lattice behavior of two metal cyanide framework compounds i.e. $KMnAg_3(CN)_6$ and $KNiAu_3(CN)_6$. We have studied the role of van der Waals dispersion and magnetic interactions on structural stability of these compounds. The behavior of these compounds under isotropic compression shows the presence of negative linear compressibility. The calculated phonon spectra, validated by inelastic neutron scattering measurements and elastic constants are used to study the negative thermal expansion behavior which is found to arise from low energy phonon modes involving the folding of A-NC-B-CN-A linkage about B atoms.

**PACS numbers:** 78.70.Nx, 63.20.-e, 65.40.–b

**Keywords**: Neutron inelastic scattering, Phonons in crystal lattice, Thermal properties of crystalline solids

## I. INTRODUCTION

Materials having open structure may show negative thermal expansion (NTE) behavior over a wide range of temperature[1-3]. The technological relevance of these compounds and their composites principally derives from their ability to withstand thermal shock without any damage. Most of these polyhedral structures are connected through their vertices[1, 3, 4]. The introduction of other anions between the polyhedral units increases the flexibility of structures through hinge like movements[5]. This type of structural flexibility has been identified in some metal organic framework structures (MOF)[5-8]. A combination of polyhedral units of metal ions coupled to organic linkages gives rise to MOF compounds which exhibit interesting lattice behavior[9-11]. The flexibility, porous nature and anisotropic bonding give rise to the interesting mechanism of negative linear compressibility (NLC), NTE as well as phase transitions in MOF's when exposed to varying temperature-pressure environments[5, 9, 11]. Different mechanisms responsible for NTE behaviour in metal inorganic and metal organic framework structures give rise to difference in magnitude of thermal expansion coefficients in these compounds[9, 12]. MOF show very large NTE coefficient and large compressibilities due to their highly anisotropic and flexible structures. The MOF exhibiting NLC along particular crystal direction are often found to show NTE along the same direction with similar mechanism[5, 9, 11].

Earlier we measured the phonon spectra of MCN (M=Ag, Au, and Cu)[10] and $ZnAu_2(CN)_4$ compounds[9]. These MCN compounds resemble one dimensional structure. These compounds show NTE along the hexagonal c-direction and positive thermal expansion behavior in other directions[9, 10]. It has been observed that occurrence of NTE in MCN compounds originated from softening of phonons induced by CN disorder in the structure[10], while in $ZnAu_2(CN)_4$ the NTE is found to be related to the phonon modes that involve bending of the -Zn-NC-Au-CN-Zn- linkage[9]. $Zn(CN)_2$ is another NTE material that exhibits CN disorder[13]. Earlier first principles calculations and Raman spectroscopic experiments suggested[14] that the soft phonons in $Zn(CN)_2$ correspond to the librational and translational modes of C≡N rigid unit. However, later another ab-initio calculations[15, 16] and high-pressure inelastic neutron scattering experiments[17] identified a broad spectrum of phonons involving low-energy acoustic modes as well as optic modes extending over a large part of the Brillouin zone comprising of translations of all the atoms and librations and distortions of $Zn(C/N)_4$ polyhedra that contribute to the NTE in this compound. In the case of $KNiAu_3(CN)_6$ and $KMnAg_3(CN)_6$ there is no such CN disorder reported in the structure[18-20]. It would be interesting to understand the behaviour of NTE compounds which may have a different origin of phonon softening. Such comparative analysis will provide a comprehensive understanding regarding the role of different interactions, disorder, as well as dimensionality on the anomalous thermal expansion behavior.

Kamali et al[18] have reported the negative linear compressibility and negative thermal expansion behaviour of $KMnAg_3(CN)_6$ using Raman spectroscopy and DFT calculations. The calculations included dispersion interactions via the vdW-DF2 approach of Langreth and Lundqvist groups. These modifications underestimated unit cell volume by 6% and c-axis by 5%. Moreover, the authors performed the DFT calculations in the non-magnetic structure without considering the effect of Hubbard onsite interactions. However, these two interactions are important in these compounds as they contain magnetic 3d atoms like Mn and Ni. Further, the thermal expansion calculations were performed by using the estimates of phonon anharmonicity from only the zone-centre modes. It is well known[21] that the phonons in the entire Brillouin zone need to be considered for calculation of the thermal expansion behavior.

We have performed inelastic neutron scattering measurements along with the ab-initio lattice dynamic calculations in the entire Brillouin zone to understand the mechanism of NTE in $KNiAu_3(CN)_6$ and $KMnAg_3(CN)_6$. Detailed analysis of eigen-vectors of phonon modes has been performed to further elaborate the nature of dynamics of different species in the compounds leading to the thermal contraction. The analysis provides some interesting physics regarding NTE in these compounds, especially the nature of anharmonic vibrations which may be associated with both NLC and NTE, and the role of magnetism.

## II. EXPERIMENTAL AND COMPUTATIONAL DETAILS

Inelastic neutron scattering measurements of the phonon density of states are performed on 2 cc volume of polycrystalline sample of $KMnAg_3(CN)_6$ in the direct-geometry cold-neutron time-of-flight time-focusing spectrometer IN6 at the Institut Laue Langevin (ILL, Grenoble, France). The spectrometer is equipped with a large detector bank covering a wide range of about 13° to 114° of scattering angle. The measurements are performed at various temperatures; namely 150, 225, 300, and 400 K, in the neutron energy gain setup and high-resolution mode, using an incident wavelength of 5.12 Å (3.12 meV). In the incoherent one-phonon approximation[22, 23], the measured scattering function $S(Q,E)$ as observed in the neutron experiments is related to the phonon density of states $g^{(n)}(E)$

$$g^{(n)}(E) = A \left\langle \frac{e^{2W(Q)}}{Q^2} \frac{E}{n(E,T) + \frac{1}{2} \pm \frac{1}{2}} S(Q,E) \right\rangle \quad (1)$$

$$g^n(E) = B \sum_k \left\{ \frac{4\pi b_k^2}{m_k} \right\} g_k(E) \quad (2)$$

where the + and − signs correspond to energy loss and gain of the neutrons respectively and $n(E,T) = [\exp(E/k_B T) - 1]^{-1}$. $A$ and $B$ are normalization constants. $b_k$, $m_k$, and $g_k(E)$ are, respectively, the neutron scattering length, mass, and partial density of states of the $k^{th}$ atom in the unit cell. The quantity between <> represents suitable average over all wave-vector transfer ($Q$) values at a given energy transfer (E). $2W(Q)$ is the Debye-Waller factor averaged over all the atoms. The weighting factors $\frac{4\pi b_k^2}{m_k}$ in the units of barns/amu for K, Mn, Ni, C, N and Ag are: 0.0501, 0.0551, 0.3152, 0.4625, 0.8221 and 0.0462 respectively. The values of neutron scattering lengths for various atoms can be found from Ref.[24].

The Vienna based ab-initio simulation package (VASP)[25-27] is used to carry out the total energy calculations based on plane-wave pseudo potential methods. The calculations are performed using projected augmented wave (PAW) formalism of Kohn-Sham density functional theory with generalized gradient approximation (GGA) for exchange correlation as given by Perdew, Becke and Ernzerhof[28-30]. K-point sampling was performed using 4×4×4 mesh Monkhorst-pack scheme with a plane wave energy cutoff of 900 eV. Various schemes allow incorporating the effect of van der Walls (vdW) interaction[31, 32] in VASP using different approximations. The phonon frequencies in the entire Brillouin zone are calculated using finite displacement method, within the direct method approach, as implemented in PHONON5.2[33]. Hellman-Feynman forces are calculated by the finite displacement of 0.03 Å.

## III. RESULTS AND DISCUSSIONS

$KAB_3(CN)_6$, where A=Mn, Ni and B=Ag, Au are metal cyanide framework compounds that crystallize in a hexagonal structure (space group: $P3_12$, 149)[34]. Their structures (**Fig. 1**) comprise of $A(CN)_6$ octahedral units where C-N is a rigid bond and connected to A through A-N bond. The nearest

A(CN)$_6$ octahedra are connected through B atoms forming A-NC-B-CN-A bonds. The K atom is placed beautifully at the center in one half of the hexagonal cell at position (2/3,1/3,1/2). The long-interconnected linkages and their dynamics could give rise to flexibility of the structure and hence anomalous lattice behavior in these materials[18-20, 34].

## A. Structure Optimization

Initializing from the experimental low-temperature structure parameters[19], we have performed complete relaxation of the KAB$_3$(CN)$_6$ unit cell using DFT with GGA exchange correlation functional[28, 29]. Although GGA based calculations are usually known to produce the structural parameters within an error of +-5%, but in case of KMnAg$_3$(CN)$_6$ the hexagonal c-parameter is highly underestimated (more than 10%). To overcome this issue we have performed structure optimization using various schemes of Dion et al[31, 32]. The introduction of van der Waals dispersion interactions through various DFT-functional schemes relaxes the structural parameters closer to the experimental values. We have tried various vdW functionals and found that optB88-vdW[35] gives the best result (TABLE I and II). The optB88 is used for further calculations. These dispersion interactions affect the structure through Ag-C bond length. These compounds contain magnetic transition metal atoms such as Mn and Ni, which have highly correlated 3d electrons. As the experimental magnetic structure is not available for these compounds, we have used simple ferromagnetic interactions among the magnetic atoms of these compounds. The ferromagnetic interactions give rise to a net magnetic moment of 4.5 and 1.8$\mu_B$ of magnetic unit cell in the Ag- and Au- compounds respectively. These magnetic interactions lower the crystal potential energy and stabilize the structure, and produce lattice parameters closer to the experimental values.

The localization of 3d orbitals of Mn and Ni is taken care through the Hubbard onsite interaction using Dudarev's approach[36]. We have taken the Hubbard U parameters for Mn and Ni atoms for metal organic frameworks[37]. Further, we have performed calculations (TABLE I and II) with varying U parameters around these values to get the best structure close to the experiments. For this purpose, an effective $U_{eff}$= U-J (5.5 for Mn and 6.7 for Ni)[37] is used to modify the pseudopotential of Mn and Ni atoms which in turn controls the binding of these atoms with their nearest neighbors.  These optimizations give rise to very good agreement between the calculated and experimental lattice parameters of these compounds.  Our calculations (at 0K) give an overestimated c-lattice parameter compared to the experiment (at 300K) by 1.7 %, which is consistent with the anisotropic NTE behavior along the c-axis[18, 19].  Moreover, our calculated (0K) volume is 2.8% lower than the experiments (100K), which is in line with the net positive volume thermal expansion in this compound[18, 19].

## B. Elastic Properties of KAB$_3$(CN)$_6$

Structural changes of the unit cell of KAB$_3$(CN)$_6$ are calculated by applying various isotropic pressures. The calculated lattice parameters as a function of applied pressure are shown in Fig. 2. The calculated pressure dependence of lattice parameters of KMnAg$_3$(CN)$_6$ shows a fair agreement with the experimental[19] values. Experimental results for KNiAu$_3$(CN)$_6$ are not available; however, both these compounds show qualitatively similar pressure dependence of lattice parameters. The c-lattice parameters of both compounds show an interesting increase with increase in applied stress and hence signify negative linear compressibility (NLC) along the hexagonal c-axis. This type of behavior is found to exist in very few compounds.

We have calculated the elastic-constant tensor (TABLE III) from stress-strain relationship obtained from finite distortions of the equilibrium lattice. The Born stability criteria[38] for a hexagonal crystal at ambient pressure are given by: $C_{44}>0$, $C_{66}>0$, $C_{11}-C_{12}>0$ and $C_{33}(C_{11}+C_{12})-2(C_{13})^2>0$. It can be seen that all four Born stability criteria for the mechanical stability of both compounds are satisfied at ambient pressure.

The calculated elastic constants are further used to determine the linear compressibility and elastic compliance matrix s=C$^{-1}$ (TABLE IV). For a hexagonal crystal[39], the linear compressibility along the crystallographic axes can be calculated from the following relations: $K_a = s_{11} + s_{12} + s_{13}$ and $K_c = s_{13} + s_{23} + s_{33}$. The compressibility (TABLE IV) along a-axis shows normal positive behavior, while along c-axis it is negative. The calculations for KMnAg$_3$(CN)$_6$ are in excellent agreement with the experimental values[19] of $K_a$ =33.2(13) TPa$^{-1}$ and $K_c$=−12.0(8) TPa$^{-1}$. The comparison between the experimental[19] and calculated pressure dependence of lattice parameters of KMnAg$_3$(CN)$_6$ is shown in **Fig. 2**, along with the calculated pressure dependence of lattice parameters of KNiAu$_3$(CN)$_6$.

The unusual compressibilities of these compounds can be understood from the calculated elastic constant tensors [TABLE III and IV]. The higher values of elastic constants for KNiAu$_3$(CN)$_6$ than that for KMnAg$_3$(CN)$_6$ implies the relatively less flexible nature of bonding in KNiAu$_3$(CN)$_6$ than in KMnAg$_3$(CN)$_6$. This causes lower NLC value in KNiAu$_3$(CN)$_6$ compared to KMnAg$_3$(CN)$_6$. The C$_{33}$ component of elastic constant tensor for both compounds is about 5 to 6 times the value of the corresponding $C_{11}$ and $C_{13}$ (or $C_{23}$). Therefore, compression or expansion along c-axis must arise from the shear stress components and not from the tensile one.

**C. Phonon Spectra of KAB$_3$(CN)$_6$**

The phonon spectra for KMnAg$_3$(CN)$_6$ (**Fig. 3**) are measured at 150 K, 225 K, 300 K and 400 K. As measurements are performed in neutron energy gain mode, we could get experimental spectra only at low energies below 45 meV at 150 K and below 70meV at higher temperatures, which are well populated at these temperatures. The spectra consist of phonon peaks around 5, 10, 14, 20 and 35 meV. As usually expected, with increasing volume and increase in temperature, all the phonon peaks are found to shift to lower energies. This indicates that the explicit anharmonicity due to increase in the thermal amplitude of atoms is much less significant in KMnAg$_3$(CN)$_6$. This behavior contrasts with that found in ZrW$_2$O$_8$, where the contribution from explicit anharmonicity is found to be significant[4, 40, 41] and on increase of temperature the peaks in the phonon spectra at low energies are found to shift to higher energies. The double peaks around 20 meV get well separated (**Fig. 3**) with increasing temperature due to different degree of softening related to atoms contributing to these modes. As discussed below, the peaks have contributions (**Figs. 4 and 5**) mainly from C and N atoms.

The calculated neutron-weighted partial and total phonon density of states of various atoms in KMnAg$_3$(CN)$_6$ along with its experimental spectrum are shown in **Fig. 4**. All the peak features of the experimental spectrum are very well reproduced by the calculations. The calculated intensity of low energy modes is underestimated compared to the experimental intensity. This might be due to the known limitations in the inelastic neutron scattering experiments of limited Q range as well as the use of incoherent approximation in the data analysis.

The total and the partial phonon density of states are shown in **Fig. 5**. This plot helps to understand the contribution of individual atomic vibrations to the total phonon spectrum of the compound and hence to identify the contributions of the various atoms at different energy ranges. In case of KAB$_3$(CN)$_6$, the C and N atoms contribute in the entire range of vibrational spectrum, which arise from the low energy translational vibrations of -CN- bond as well as higher energy internal stretching vibrations. All other atoms like K, A (Mn, Ni) and B(Ag, Au) contribute in the low energy range upto 65meV. The maximum contribution of B atoms can be seen in low energy range around 5 meV and might give rise to interesting dynamical properties. This indicates that contributions due to atomic vibrations of B(Ag, Au) atoms mainly arise from the transverse vibrations across the A-NC-B-CN-A bond chains. The phonon peaks above 20 meV in KMnAg$_3$(CN)$_6$ are well separated in comparison to that in KNiAu$_3$(CN)$_6$. The contribution due to various atoms in the energy range of 20-55 meV of the phonon spectrum in KNiAu$_3$(CN)$_6$ is found to shift to 20-65 meV in KMnAg$_3$(CN)$_6$.

We have calculated the phonon dispersion relation of KAB$_3$(CN)$_6$ compounds over the entire Brillioun zone, which along various high symmetry directions in the Brillioun zone are shown in **Fig. 6**. The slope of acoustic phonon branches near the zone center signifies the elastic behavior of the compounds. There are four regions of phonon dispersion curves with some band gaps in between. The band gaps as seen in dispersion relation (**Fig. 6**) matches very well with that seen in the calculated partial density of states (**Fig. 5**). The stretching vibrations of -CN- bonds at about 270 meV are almost dispersion less due to the highly rigid nature of these triple bonds, and are not shown in Fig. 6.

The calculated zone-centre frequencies and their irreducible representations for both compounds are classified in Table V. These first-principles calculated data would be useful in the mode assignments of Raman and infra-red measurements as and when available.

**D. Thermal Expansion Behaviour**

Thermal expansion coefficients of a solid are related to the Grüneisen parameters and elastic properties as discussed below. The mode Grüneisen parameter of phonon of energy $E_{q,j}$ (q is the phonon wave-vector, and j (=1, 3n) is the mode index, n being the number of atoms in the primitive unit cell) is given as[42],

$$\Gamma_l(E_{q,j}) = -\left(\frac{\partial lnE_{q,j}}{\partial lnl}\right)_{l'}; \quad l,l' = a,b,c \; \& \; l \neq l' \quad (3)$$

Thermal expansion calculation is done[21] using pressure dependence of phonon frequencies in the entire Brillouin zone within the quasi-harmonic approximation. In the framework of quasiharmonic approximation, the contribution of $j^{th}$ phonon mode (of energy $E_{qj}$) at point q in the Brillouin zone to the thermal expansion coefficient, is given by the following relation[42]:

$$\alpha_l(T) = \frac{1}{V_0}\sum_{q,j} C_{l'}(q,j,T)\left[s_{l1}\Gamma_a + s_{l2}\Gamma_b + s_{l3}\Gamma_c\right], \quad l,l' = a,b,c \; \& \; l \neq l' \quad (4)$$

Where $s_{ij}$ are elements of elastic compliances matrix, $s=C^{-1}$ (TABLE IV) at constant temperature T=0 K, $V_0$ is volume at 0K and $C_{l'}(q,j,T)$ is the specific heat at constant strain due to $j^{th}$ phonon mode at point **q** in the Brillouin zone. This implies for a hexagonal system (a=b), the coefficient of volume thermal expansion is given by:

$$\alpha_V = (2\alpha_a + \alpha_c) \qquad (5)$$

The anisotropic pressure dependence of phonon spectrum is calculated to obtain the anisotropic Grüneisen parameters. For this purpose, the lattice parameters corresponding to the relaxed unit cell at a pressure of 5 Kbar are used. The anisotropic Grüneisen parameters are calculated by changing the lattice constant "a" corresponding to 5 Kbar and keeping the "c" parameter constant, and vice versa, where temperature is held constant at zero. During these calculations, the symmetry of the unit cell is kept invariant on introduction of strain. The calculated anisotropic mode Grüneisen parameters as a function of phonon energy, as averaged over all the phonons in the Brillouin zone, are shown in **Fig. 7**. The $\Gamma_a$ for both compounds shows small positive values in the entire phonon energy range except below 4 meV. However, the $\Gamma_c$ shows negative values upto 45 meV. The magnitude of negative $\Gamma_c$ is larger for $KMnAg_3(CN)_6$ than that for $KNiAu_3(CN)_6$ which could give rise to higher coefficient of thermal expansion along c-axis in $KMnAg_3(CN)_6$.

The calculated linear thermal expansion behaviour for both compounds is plotted as a function of temperature in **Fig. 8**. The c-axis of both compounds shows negative thermal expansion while a-axis shows positive thermal expansion in the entire temperature range upto 500 K. The overall volume thermal expansion coefficient is positive in these compounds. The linear $\alpha_l$ are larger in $KMnAg_3(CN)_6$ than that in $KNiAu_3(CN)_6$. The average linear thermal coefficients extracted from the diffraction data[19] of $KMnAg_3(CN)_6$ in the temperature range from 100 K to 300 K are $\alpha_a = +61(2) \times 10^{-6}$ K$^{-1}$ and $\alpha_c = -60(3) \times 10^{-6}$ K$^{-1}$, while the corresponding values for $KNiAu_3(CN)_6$ as obtained from experimental data[20] in the range of temperature from 100 to 395 K are $\alpha_a = +59.7$ to $+58.7 \times 10^{-6}$ K$^{-1}$ and $\alpha_c = -42.2$ to $-42.7 \times 10^{-6}$ K$^{-1}$. The calculated $\alpha_l$ values at 300 K are:

$KMnAg_3(CN)_6$: $\alpha_a= 60 \times 10^{-6}$K$^{-1}$, $\alpha_c= -47 \times 10^{-6}$K$^{-1}$, $\alpha_V= 85 \times 10^{-6}$K$^{-1}$

$KNiAu_3(CN)_6$: $\alpha_a= 45 \times 10^{-6}$K$^{-1}$, $\alpha_c= -32 \times 10^{-6}$K$^{-1}$, $\alpha_V= 62 \times 10^{-6}$K$^{-1}$

The overall calculated lattice parameters of these compounds as a function of temperature are compared with their experimental values[19, 20] in **Fig. 9**. The calculated and the experimental data are in accord with each other. The calculated contribution of phonon of energy E to the linear thermal expansion coefficients at 300 K is shown in **Fig. 10**. It can be seen that the phonon modes which contribute to positive expansion coefficients along a-axis also give rise to negative expansion coefficient along c-axis of the hexagonal unit cell. The phonon modes below 10 meV give larger negative $\alpha_c$ value of $-8 \times 10^{-6}$ K$^{-1}$ for KMnAg$_3$(CN)$_6$ in comparison to that of $-5 \times 10^{-6}$ K$^{-1}$ in KNiAu$_3$(CN)$_6$. The modes around 20 meV give a larger contribution to $\alpha_l$ in KNiAu$_3$(CN)$_6$ than that in KMnAg$_3$(CN)$_6$.

The calculated contribution of phonon of energy E to the mean-square displacements (u$^2$) of various atoms at 300 K for KMnAg$_3$(CN)$_6$ and KNiAu$_3$(CN)$_6$ is shown in **Fig. 11.** It can be seen that for B(Ag, Au) atoms the low energy modes around 5 meV contribute highly to the mean square displacement (u$^2$). Further, the u$^2$ values of Ag atoms in KMnAg$_3$(CN)$_6$ are much higher (**Fig. 11)** than those of Au in KNiAu$_3$(CN)$_6$, which could arise from comparatively lower mass of Ag atoms. This would give rise to more flexibility to A-NC-B-CN-A linkages in KMnAg$_3$(CN)$_6$ and may be responsible for higher coefficients of negative thermal expansion in KMnAg$_3$(CN)$_6$ in comparison to KNiAu$_3$(CN)$_6$. For low energies around 5 meV, the A(=Mn, Ni) atoms have a low vibrational amplitude in comparison to C atoms indicating that low energy phonon modes mainly involve rotation of AC$_6$. However, detailed mode analysis is required to microscopically understand the phenomenon of the anomalous thermal expansion behavior.

In order to understand the microscopic mechanism responsible for negative thermal expansion behavior along the c-axis, we have analyzed the specific phonon modes of low energy with negative α$_c$. The linear thermal expansion coefficient, due to the Γ-point mode of energy 1.42 THz for KMnAg$_3$(CN)$_6$ and 1.59THz for KNiAu$_3$(CN)$_6$ (assuming it as an Einstein mode with one degree of freedom), along various axes is: α$_a$ =6.0×10$^{-6}$K$^{-1}$, α$_c$ =−3.9×10$^{-6}$ K$^{-1}$ and α$_a$ =3.0×10$^{-6}$ K$^{-1}$, α$_c$ =−2.5× 10$^{-6}$ K$^{-1}$, respectively. This mode involves (**Fig. 12**) partially transverse vibrations of C atoms about B(Ag,Au) atoms and hence gives rise to folding of A-NC-B-CN-A linkages which lie along the c-axis. Therefore, this mode produces contraction along the c-axis. However, as folding dynamics of B atoms happens in ab-plane, it may lead to expansion in ab-plane of the unit cell. The lower value of $\alpha_l$ for KNiAu$_3$(CN)$_6$ than that of KMnAg$_3$(CN)$_6$ for this mode may arise from the lower values of the mode Grüneisen parameters in the Au-compound.

The other Γ-point mode with energy 1.86 THz for KMnAg$_3$(CN)$_6$ and 1.98 THz for KNiAu$_3$(CN)$_6$ gives rise to linear $\alpha_l$ values as: $\alpha_a$ =11.6×10$^{-6}$K$^{-1}$, $\alpha_c$ =-9.2×10$^{-6}$K$^{-1}$ and $\alpha_a$ =7.6×10$^{-6}$K$^{-1}$, $\alpha_c$ =−7.2×10$^{-6}$K$^{-1}$ for these compounds, respectively. This mode (**Fig. 12**) also involves the transverse chain dynamics of C atoms, however; there is an equal and opposite movement of two C atoms on both sides of B atoms. This gives large bending of A-NC-B-CN-A linkage about B atoms and hence contracts the c-axis and expands the a-axis. Therefore, the mechanism of NTE along c-axis arises from the bending of A-NC-B-CN-A linkage about B(Ag, Au) atoms.

IV. **Conclusions**

Inelastic neutron scattering and ab-initio calculations of phonon spectra are performed to study thermodynamic behavior of KMnAg$_3$(CN)$_6$ and KNiAu$_3$(CN)$_6$. The compounds are found to have strong vander Waals dispersion interactions along with ferromagnetic interactions among the magnetic atoms (Mn, Ni) with net magnetic moment of 4.5 $\mu_B$ and 1.8$\mu_B$ respectively. Pressure dependent DFT calculations show a negative linear compressibility along the c-axis of these compounds. This NLC also gives rise to NTE along c-axis which can be derived from the anisotropy in elastic constants and Grüneisen parameters. The NTE along the c-axis arises from the low energy modes involving folding of A-NC-B-CN-A chain due to transverse oscillations of C atoms about B(Ag,Au) atoms.

**ACKNOWLEDGEMENT**

The *Institut Laue-Langevin* (ILL, Grenoble, France) is thanked for providing beam time on the IN6 spectrometer to perform the variable-temperature inelastic neutron scattering measurements reported in this work.

TABLE I: The experimental[19] and calculated lattice parameters of $KMnAg_3(CN)_6$ using various *ab-initio* DFT schemes. The DFT calculations are performed using GGA exchange correlation and optB88-vdW functional for van der Waals interaction. The Hubbard onsite interactions (U) are taken cared using Drudav approach for 3d electrons of Mn atom.

| | a (Å) | c (Å) | V(Å$^3$) | μ (μ$_B$) |
|---|---|---|---|---|
| Expt.(100K) | 6.8214 | 8.2353 | 331.86 | - |
| DFT-GGA | 7.12319 | 7.30216 | 320.87 | - |
| GGA+Van der Waals Dispersion Interaction | | | | |
| optB88-vdW (BO) | 6.70657 | 7.65803 | 298.30 | - |
| DFT-D2 (CR) | 6.76180 | 7.55356 | 299.09 | - |
| optB86b-vdW (MK) | 6.66415 | 7.63252 | 293.55 | - |
| optPBE-vdW(OR) | 6.87528 | 7.57714 | 310.18 | - |
| revPBE-vdW(RE) | 7.13084 | 7.50089 | 330.31 | - |
| GGA+Van der Waals Dispersion Interaction(optB88-vdW) + Hubbard Onsite Interaction (U)+ Magnetic (FM) | | | | |
| GGA+vdW+ FM | 6.70593 | 7.66815 | 298.63 | 0.941 |
| GGA+vdW+U5.5+FM | 6.66638 | 8.37780 | 322.43 | 4.547 |

TABLE II: The experimental[20] and calculated lattice parameters of $KNiAu_3(CN)_6$ using various *ab-initio* DFT schemes. The DFT calculations are performed using GGA exchange correlation and optB88-vdW functional for van der Waals interaction. The Hubbard onsite interactions (U) are taken cared using Drudav approach for 3d electrons of Ni atom.

| | a | c | V | M |
|---|---|---|---|---|
| Expt.(100K) | 6.7531 | 7.7517 | 306.15 | |
| DFT-GGA | 7.08258 | 7.48489 | 325.16 | |
| GGA+Van der Waals Dispersion Interaction | | | | |
| optB88-vdW (BO) | 6.66846 | 7.80105 | 300.42 | - |
| DFT-D2 (CR) | 7.03724 | 7.41792 | 318.34 | - |
| optB86b-vdW (MK) | 6.59286 | 7.80952 | 293.97 | - |
| optPBE-vdW(OR) | 6.84027 | 7.72666 | 313.09 | - |
| revPBE-vdW(RE) | 7.13174 | 7.61926 | 335.61 | - |
| GGA+Van der Waals Dispersion Interaction (optB88-vdW) + Hubbard Onsite Interaction (U)+ Magnetic (FM) | | | | |
| GGA+vdW+U6.70+FM | 6.70986 | 7.89750 | 307.93 | 1.801 |

TABLE III. The calculated elastic constant for $KMnAg_3(CN)_6$ and $KNiAu_3(CN)_6$ in GPa units.

| Elastic constant | $KMnAg_3(CN)_6$ | $KNiAu_3(CN)_6$ |
|---|---|---|
| $C_{11}$ | 23.9 | 33.0 |
| $C_{33}$ | 130.7 | 158.2 |
| $C_{44}$ | 14.6 | 26.6 |
| $C_{12}$ | 16.1 | 21.2 |
| $C_{13}$ | 37.2 | 52.0 |
| $C_{66}$ | 3.9 | 5.9 |

TABLE IV. The calculated elastic compliances, $s=C^{-1}$ for $KMnAg_3(CN)_6$ and $KNiAu_3(CN)_6$ respectively in $GPa^{-1}$ unit. The quantities in the brackets along with the calculated values for $KMnAg_3(CN)_6$ are experimental data.

| Compliance | $KMnAg_3(CN)_6$ | $KNiAu_3(CN)_6$ |
|---|---|---|
| $s_{11}$ ($GPa^{-1}$) | 0.0906 | 0.0675 |
| $s_{33}$ ($GPa^{-1}$) | 0.0163 | 0.0171 |
| $s_{44}$ ($GPa^{-1}$) | 0.0687 | 0.0376 |
| $s_{12}$ ($GPa^{-1}$) | -0.0372 | -0.0174 |
| $s_{13}$ ($GPa^{-1}$) | -0.0152 | -0.0165 |
| $K_a$ ($GPa^{-1}$) | 0.0382 (0.0332(13)) | 0.0336 |
| $K_c$ ($GPa^{-1}$) | -0.0141(-0.0120(8)) | -0.0159 |

TABLE V. The calculated phonon frequencies classified in transverse and longitudinal phonon modes and corresponding irreducible representations for $KMnAg_3(CN)_6$ and $KNiAu_3(CN)_6$, respectively (1 THz=4.136 meV).

| colspan="4" | $KMnAg_3(CN)_6$ | | | | $KNiAu_3(CN)_6$ | | |
|---|---|---|---|---|---|---|---|
| Type | Multi. | Omega(THz) | Irreducible Representations | Type | Multi. | Omega(THz) | Irreducible Representations |
| TO | 1 | 0.152 | A2(I) | TO | 1 | 0.033 | A2(I) |
| TO | 1 | 1.104 | A2(I) | TO | 1 | 1.130 | A2(I) |
| TO | 1 | 1.385 | E(RI) | TO | 1 | 1.336 | E(RI) |
| TO | 1 | 1.420 | E(RI) | TO | 1 | 1.587 | E(RI) |
| TO | 1 | 1.807 | A1(R) | TO | 1 | 1.857 | A1(R) |
| TO | 1 | 1.863 | A2(I) | TO | 1 | 1.988 | A2(I) |
| TO | 1 | 2.463 | E(RI) | TO | 1 | 2.767 | E(RI) |
| TO | 1 | 3.401 | E(RI) | TO | 1 | 3.683 | E(RI) |
| TO | 1 | 3.917 | E(RI) | TO | 1 | 4.407 | A2(I) |
| TO | 1 | 3.959 | A2(I) | TO | 1 | 4.408 | E(RI) |
| TO | 1 | 4.431 | A1(R) | TO | 1 | 5.792 | A1(R) |
| TO | 1 | 5.227 | E(RI) | TO | 1 | 6.271 | E(RI) |
| TO | 1 | 5.779 | A1(R) | TO | 1 | 6.307 | A2(I) |
| TO | 1 | 5.834 | E(RI) | TO | 1 | 6.647 | E(RI) |
| TO | 1 | 5.938 | A2(I) | TO | 1 | 6.697 | A1(R) |
| TO | 1 | 6.333 | E(RI) | TO | 1 | 7.442 | E(RI) |
| TO | 1 | 6.798 | A2(I) | TO | 1 | 7.756 | A2(I) |
| TO | 1 | 7.953 | E(RI) | TO | 1 | 9.444 | E(RI) |
| TO | 1 | 8.356 | A2(I) | TO | 1 | 9.921 | A2(I) |
| TO | 1 | 8.838 | E(RI) | TO | 1 | 10.620 | A1(R) |
| TO | 1 | 8.987 | A1(R) | TO | 1 | 10.689 | E(RI) |
| TO | 1 | 9.374 | E(RI) | TO | 1 | 12.466 | E(RI) |
| TO | 1 | 9.755 | A2(I) | TO | 1 | 12.631 | A2(I) |
| TO | 1 | 10.057 | E(RI) | TO | 1 | 12.785 | E(RI) |
| TO | 1 | 10.286 | A1(R) | TO | 1 | 12.966 | A1(R) |
| TO | 1 | 11.619 | E(RI) | TO | 1 | 13.742 | A2(I) |
| TO | 1 | 12.912 | A2(I) | TO | 1 | 13.859 | E(RI) |
| TO | 1 | 12.931 | A1(R) | TO | 1 | 14.317 | E(RI) |
| TO | 1 | 12.945 | E(RI) | TO | 1 | 15.411 | A1(R) |
| TO | 1 | 65.850 | E(RI) | TO | 1 | 65.938 | E(RI) |
| TO | 1 | 65.852 | A2(I) | TO | 1 | 65.961 | A2(I) |
| TO | 1 | 66.100 | E(RI) | TO | 1 | 66.488 | E(RI) |
| TO | 1 | 66.278 | A1(R) | TO | 1 | 66.807 | A1(R) |
| LO | 1 | 1.385 | mixture of 17E(RI) | LO | 1 | 1.336 | mixture of 16E(RI) |
| LO | 1 | 1.425 | mixture of 17E(RI) | LO | 1 | 1.590 | mixture of 16E(RI) |
| LO | 1 | 2.470 | mixture of 17E(RI) | LO | 1 | 2.801 | mixture of 16E(RI) |
| LO | 1 | 3.401 | mixture of 17E(RI) | LO | 1 | 3.684 | mixture of 16E(RI) |
| LO | 1 | 3.950 | mixture of 17E(RI) | LO | 1 | 4.461 | mixture of 16E(RI) |
| LO | 1 | 5.241 | mixture of 17E(RI) | LO | 1 | 6.272 | mixture of 16E(RI) |
| LO | 1 | 5.885 | mixture of 17E(RI) | LO | 1 | 6.647 | mixture of 16E(RI) |
| LO | 1 | 6.421 | mixture of 17E(RI) | LO | 1 | 7.914 | mixture of 16E(RI) |
| LO | 1 | 7.953 | mixture of 17E(RI) | LO | 1 | 9.445 | mixture of 16E(RI) |
| LO | 1 | 8.838 | mixture of 17E(RI) | LO | 1 | 10.689 | mixture of 16E(RI) |
| LO | 1 | 9.376 | mixture of 17E(RI) | LO | 1 | 12.466 | mixture of 16E(RI) |
| LO | 1 | 10.060 | mixture of 17E(RI) | LO | 1 | 12.785 | mixture of 16E(RI) |
| LO | 1 | 11.619 | mixture of 17E(RI) | LO | 1 | 13.898 | mixture of 16E(RI) |
| LO | 1 | 12.955 | mixture of 17E(RI) | LO | 1 | 14.319 | mixture of 16E(RI) |
| LO | 1 | 65.862 | mixture of 17E(RI) | LO | 1 | 65.982 | mixture of 16E(RI) |
| LO | 1 | 66.100 | mixture of 17E(RI) | LO | 1 | 66.488 | mixture of 16E(RI) |

FIG 1.(Color Online) Structure of $KAB_3(CN)_6$, where A(Blue)=Mn, Ni and B(Green)=Ag, Au. Here, black, red and violet spheres represent C, N and K atoms, respectively.

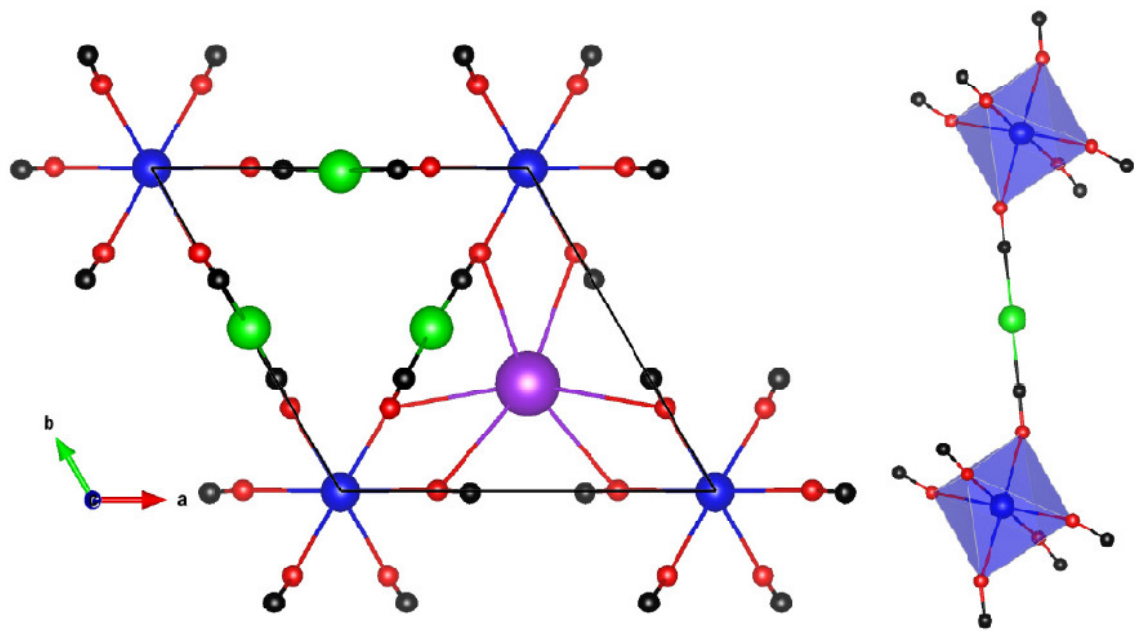

FIG 2.(Color Online) (a) The experimental[19] and calculated pressure dependence of lattice parameters of $KMnAg_3(CN)_6$.(b) The calculated pressure dependence of lattice parameters of $KNiAu_3(CN)_6$.

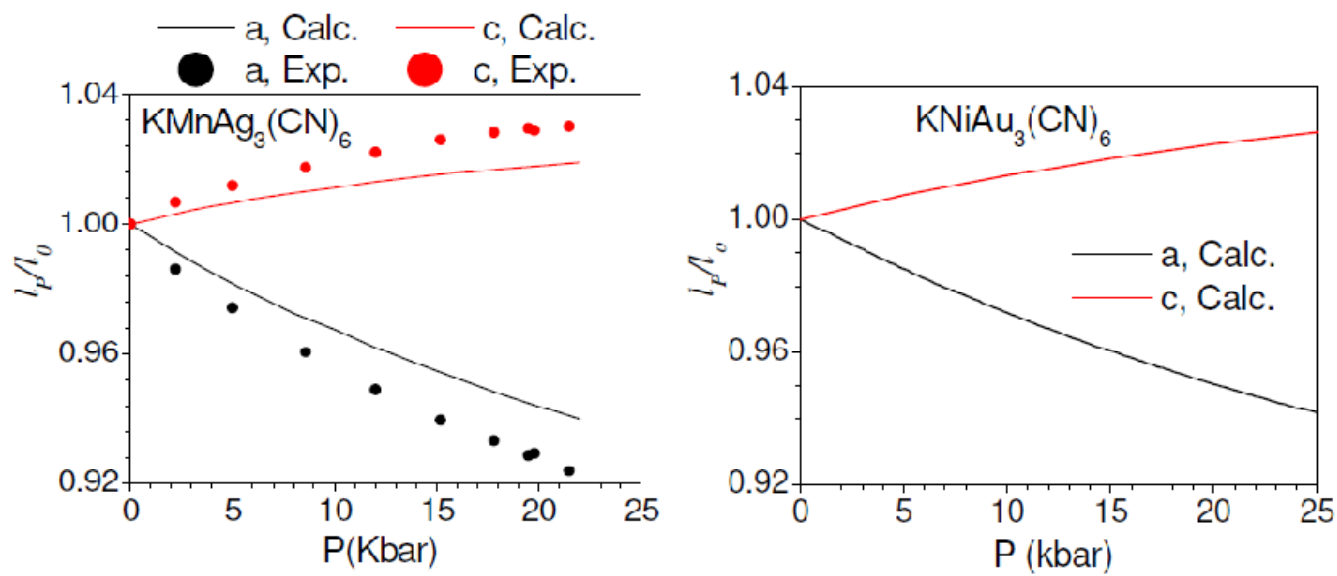

FIG 3.(Color Online) The experimentally measured temperature dependent inelastic neutron scattering(phonon) spectra of KMnAg$_3$(CN)$_6$ using the IN6 spectrometer. The experimental data at various temperatures are shifted vertically for clarity.

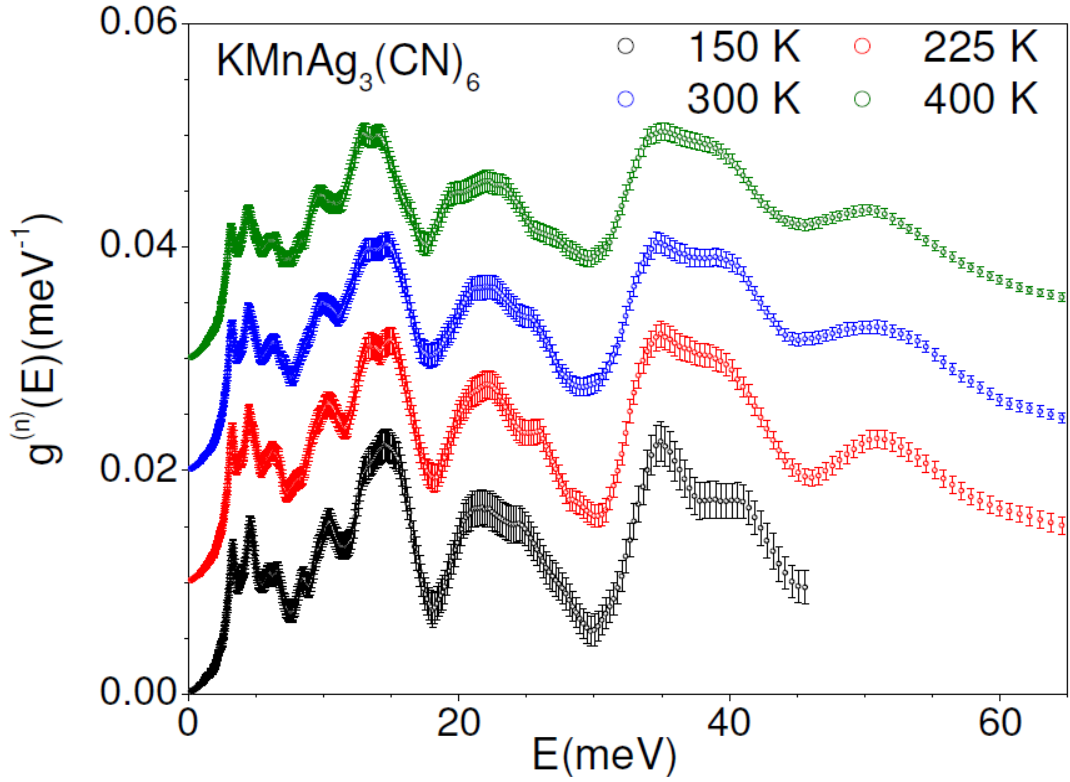

FIG4.(Color Online) (a) The experimentally measured, using IN6, and calculated total and partial neutron weighted phonon densities of states of various atoms in KMnAg$_3$(CN)$_6$. (b) The calculated neutron weighted partial and total phonon densities of states of various atoms in KNiAu$_3$(CN)$_6$, respectively.

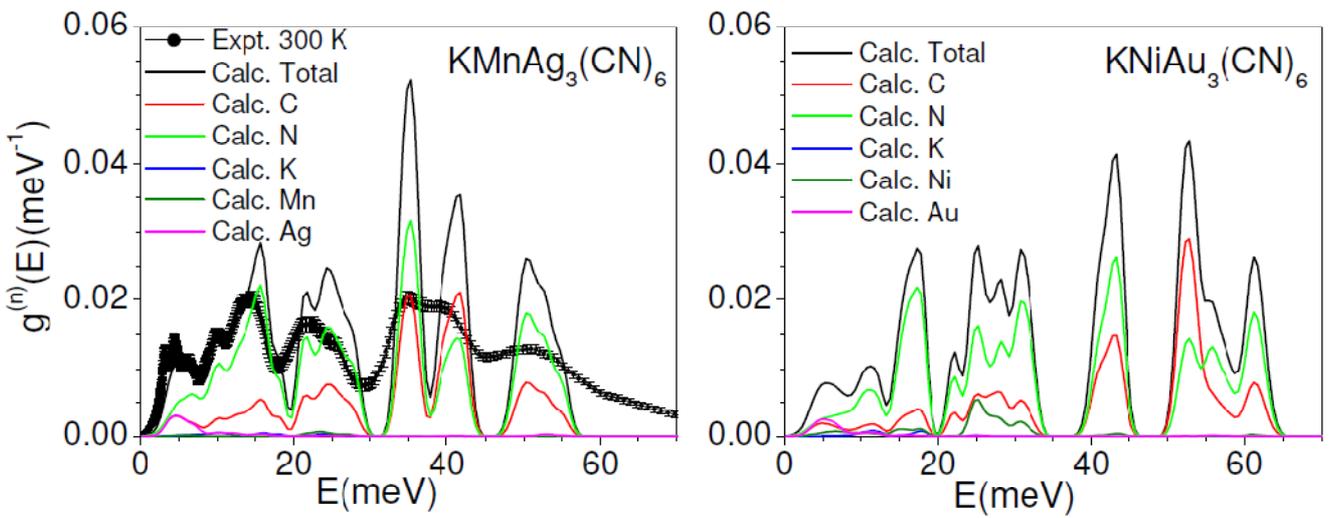

FIG 5.(Color Online) The calculated total and the partial phonon densities of states of various atoms in KMnAg$_3$(CN)$_6$ and KNiAu$_3$(CN)$_6$, respectively.

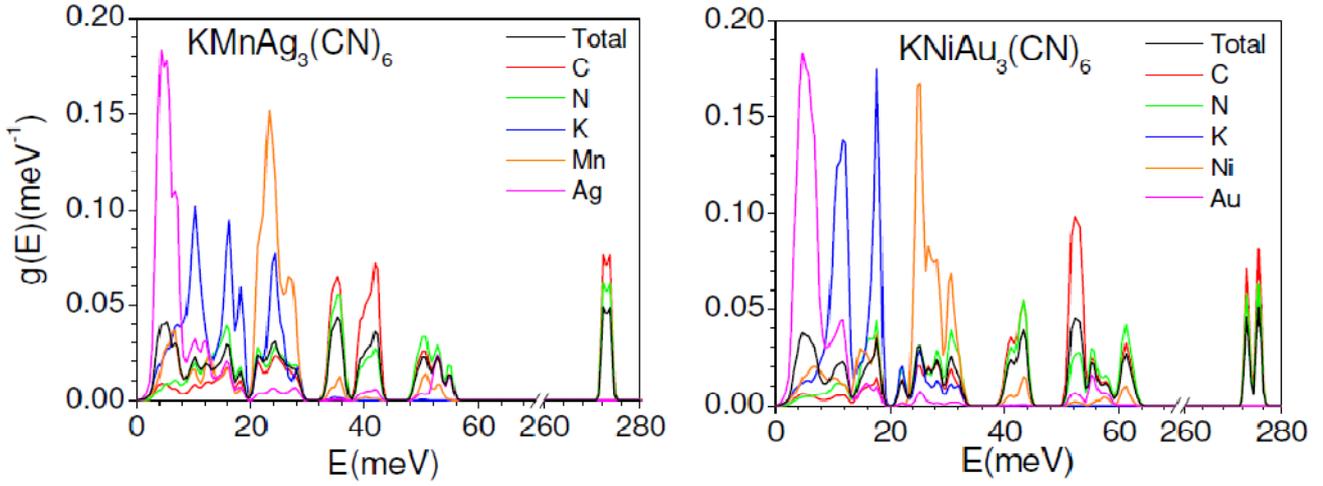

FIG 6. The calculated phonon dispersion curve along the high symmetry directions in the Brillouin zone of KMnAg$_3$(CN)$_6$ and KNiAu$_3$(CN)$_6$, respectively. The stretching vibrations of -CN- bonds at about 270 meV (not shown in the figure) are almost dispersion less due to the highly rigid nature of these triple bonds, and are not shown here.

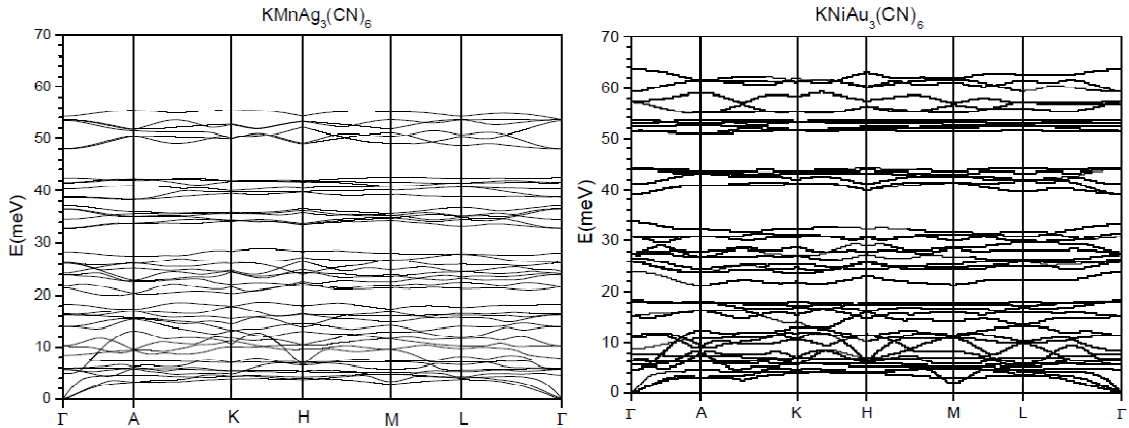

FIG7. (Color Online) The calculated anisotropic Grüneisen parameters as a function of phonon energy, as averaged over all the phonons in the Brillouin zone, in $KMnAg_3(CN)_6$ and $KNiAu_3(CN)_6$, respectively.

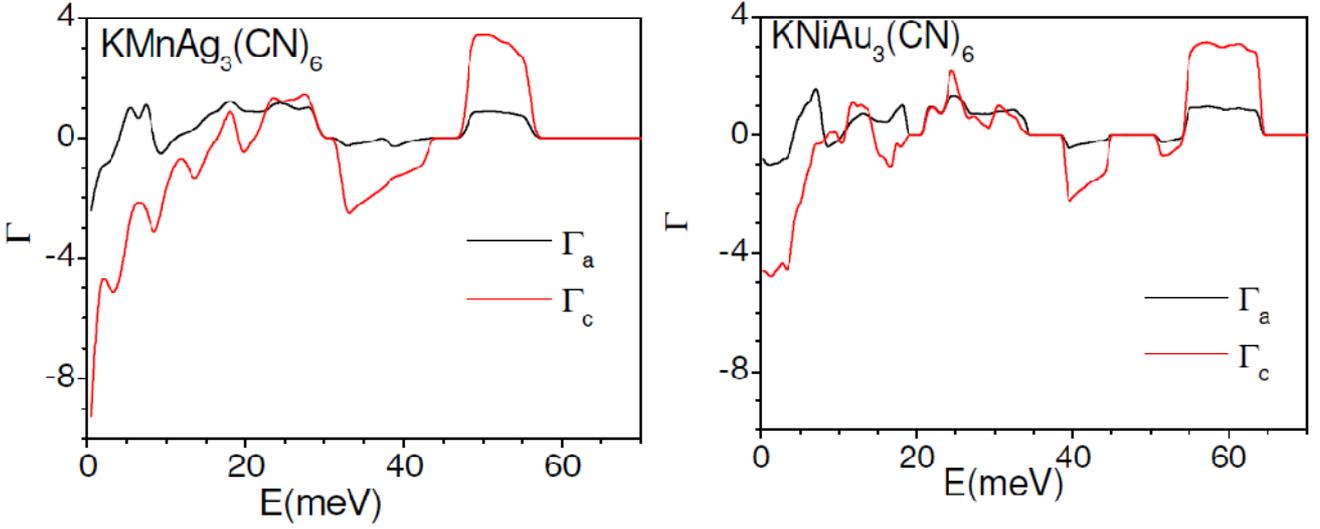

FIG 8. (Color Online) The calculated temperature-dependent linear and volume thermal expansion coefficients of $KMnAg_3(CN)_6$ and $KNiAu_3(CN)_6$ respectively.

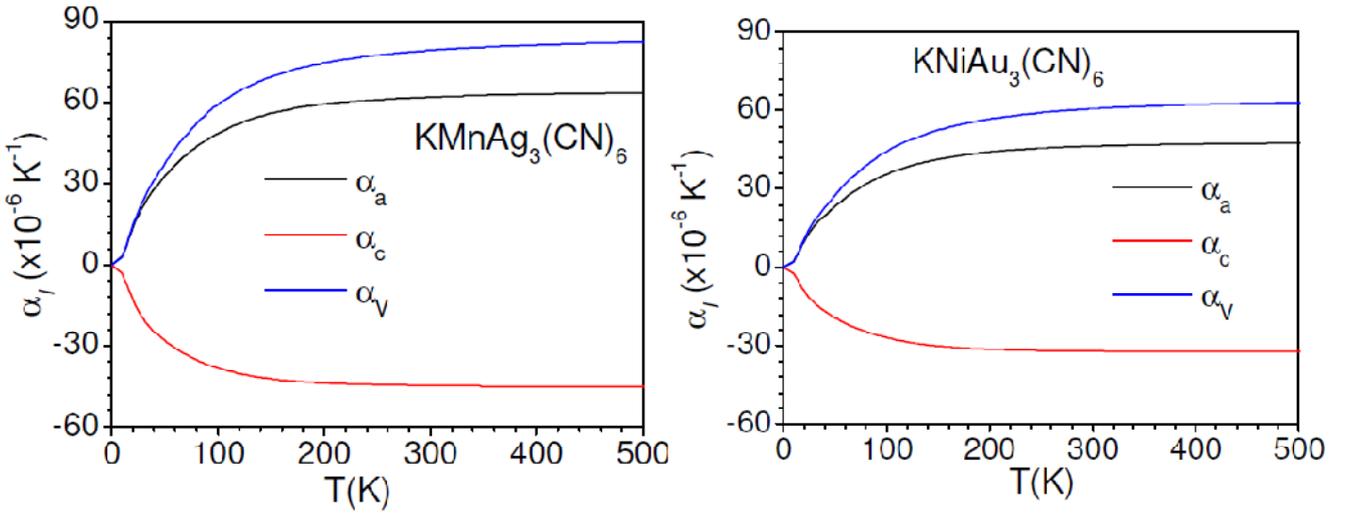

FIG 9.(Color Online) The experimental[19, 20] and calculated temperature dependence of lattice parameters in KMnAg$_3$(CN)$_6$ and KNiAu$_3$(CN)$_6$, respectively.

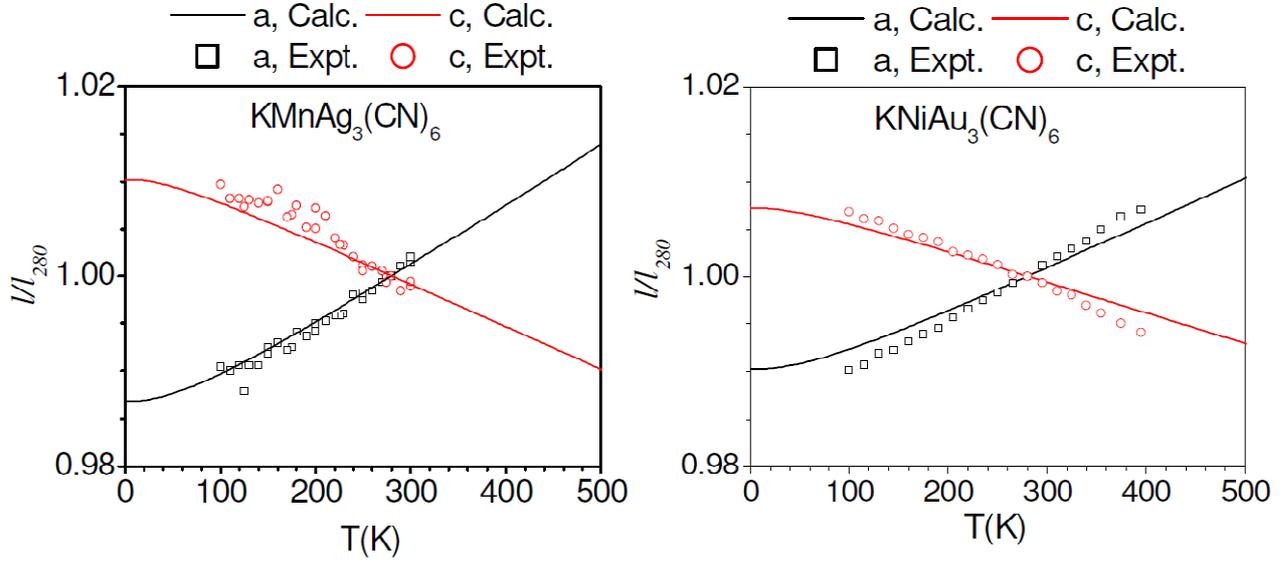

FIG 10. (Color Online) The calculated contribution of phonons of energy E to the linear thermal expansion coefficients at 300 K in KMnAg$_3$(CN)$_6$ and KNiAu$_3$(CN)$_6$.

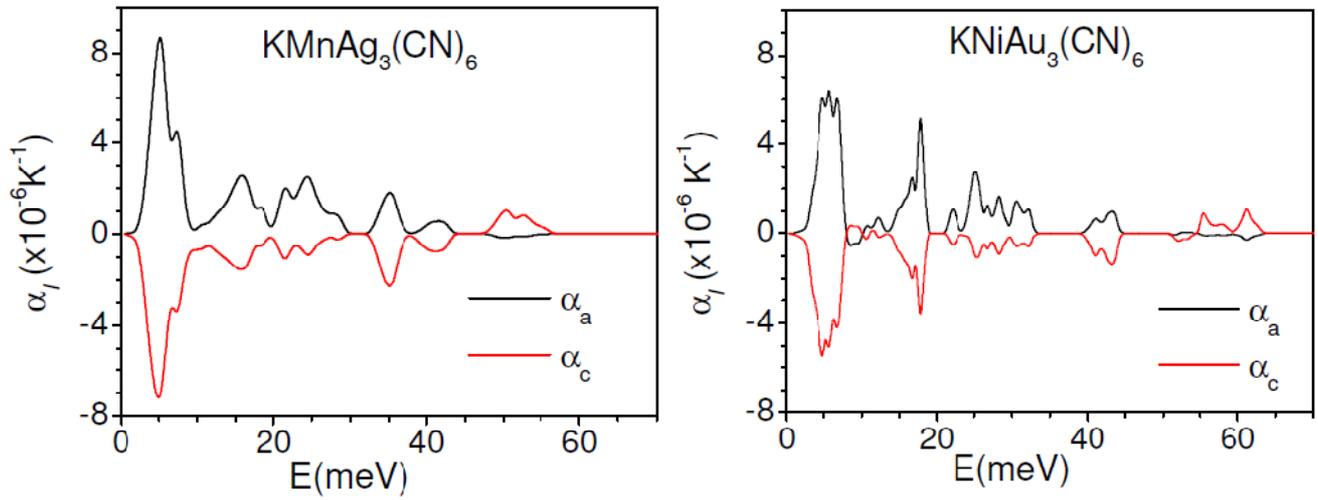

FIG 11.(Color Online) The calculated contribution of phonons of energy E to the mean squared displacements ($u^2$) at 300 K in KMnAg$_3$(CN)$_6$ and KNiAu$_3$(CN)$_6$.

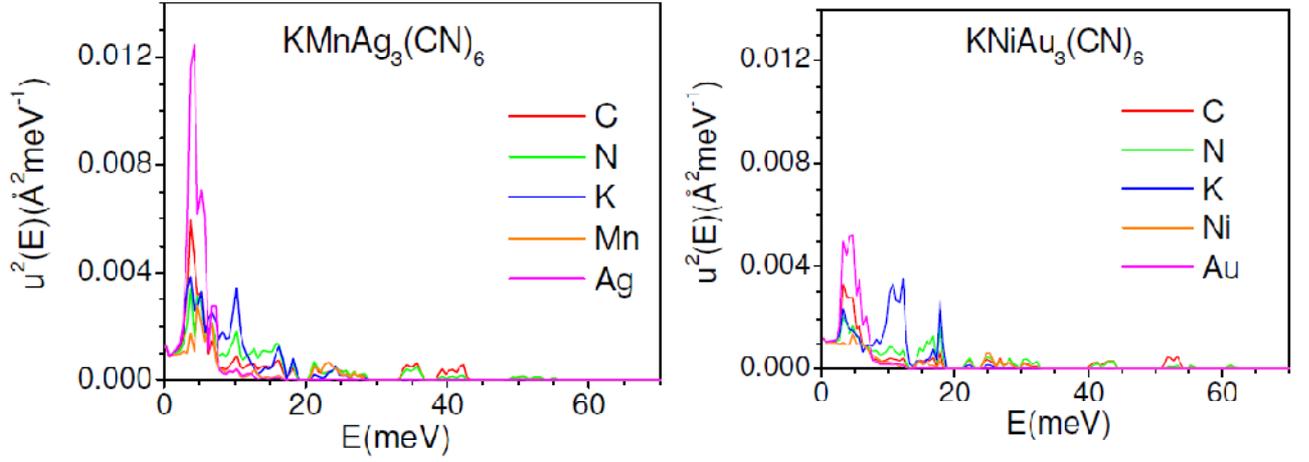

FIG 12. (Color Online) The calculated eigenvector of selected zone centre (Γ- Point) optic phonon modes giving rise to negative thermal expansion along hexagonal c-axis in KMnAg$_3$(CN)$_6$ and KNiAu$_3$(CN)$_6$, respectively. The values of linear thermal expansion coefficients ($\alpha_a$ and $\alpha_c$) are at 300 K and are in the units of $10^{-6}$ K$^{-1}$. Key: A(Mn, Ni), blue sphere; B(Ag, Au), Green sphere; C, black sphere; N, red sphere; K, violet sphere .

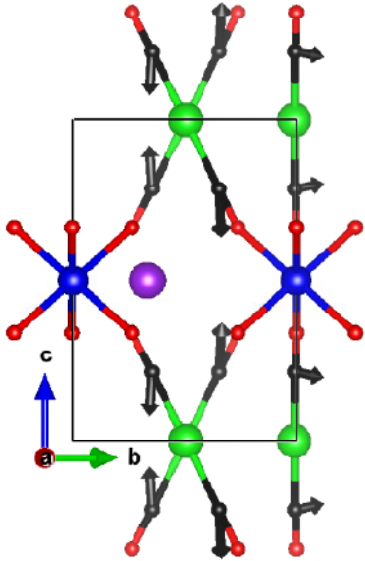 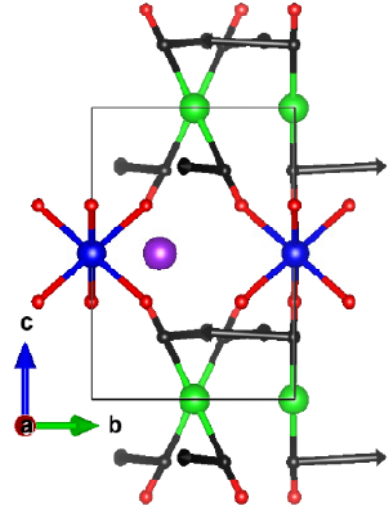

**KMnAg$_3$(CN)$_6$: 1.42 THz**
$\Gamma_a$=2.1, $\Gamma_c$=-1.7, $\alpha_a$=6.0, $\alpha_c$=-3.9

**KNiAu$_3$(CN)$_6$: 1.59 THz**
$\Gamma_a$=0.7, $\Gamma_c$=-1.9, $\alpha_a$=3.0, $\alpha_c$=-2.5

**KMnAg$_3$(CN)$_6$: 1.86 THz**
$\Gamma_a$=2.8, $\Gamma_c$=-8.0, $\alpha_a$=11.6, $\alpha_c$=-9.2

**KNiAu$_3$(CN)$_6$: 1.98 THz**
$\Gamma_a$=0.9, $\Gamma_c$=-7.8, $\alpha_a$=7.6, $\alpha_c$=-7.2